\documentclass[12pt]{article}

\usepackage{graphicx}
\usepackage[switch, modulo]{lineno}

\usepackage{times}

\newenvironment{sciabstract}{%
\begin{quote} \bf}
{\end{quote}}

\newcommand{\beginsupplement}{%
        \setcounter{table}{0}
        \renewcommand{\thetable}{S\arabic{table}}%
        \setcounter{figure}{0}
        \renewcommand{\thefigure}{S\arabic{figure}}%
     }

\newcounter{lastnote}

\begin{document}
\begin{center}
{\LARGE Decline of long-range temporal correlations in the human brain during sustained wakefulness}\\
\vspace{0.8cm}
{\large Christian Meisel$^{1,2\ast}$, Kimberlyn Bailey$^{1}$, Peter Achermann$^{3}$ and Dietmar Plenz$^{1}$}\\
\end{center}


\date{}



\baselineskip24pt



\newpage
\begin{sciabstract}
Sleep is crucial for daytime functioning, cognitive performance and general well-being. These aspects of daily life are known to be impaired after extended wake, yet, the underlying neuronal correlates have been difficult to identify.  Accumulating evidence suggests that normal functioning of the brain is characterized by long-range temporal correlations (LRTCs) in cortex, which are supportive for decision-making and working memory tasks. 

Here we assess LRTCs in resting state human EEG data during a 40-hour sleep deprivation experiment by evaluating the decay in autocorrelation and the scaling exponent of the detrended fluctuation analysis from EEG amplitude fluctuations. We find with both measures that LRTCs decline as sleep deprivation progresses. This decline becomes evident when taking changes in signal power into appropriate consideration. 

Our results demonstrate the importance of sleep to maintain LRTCs in the human brain. In complex networks, LRTCs naturally emerge in the vicinity of a critical state. The observation of declining LRTCs during wake thus provides additional support for our hypothesis that sleep reorganizes cortical networks towards critical dynamics for optimal functioning.

\end{sciabstract}


\newpage
\section*{Introduction}

Sleep is essential for daytime functioning and well-being.
Without sleep optimal brain functioning such as responsiveness to stimuli, information processing, or learning is impaired \cite{VanDongen2003,Banks2007,Mignot2008,Killgore2010}.
The neuronal correlates and mechanisms by which sleep improves, or, conversely, by which the lack of sleep impairs cognitive function and information processing in cortical networks, are largely still not understood. 

An essential ingredient for information processing is thought to be the ability of neural circuits to to integrate information over extended periods of time. For example, in decision-making and working memory tasks \cite{Kiebel2008,Friston2012,Chaudhuri2015,Kringelbach2015}, this ability may increase the signal-to-noise ratio and to maintain some memory of past activity. The network dynamics typically observed with this ability are characterized by slowly decaying autocorrelations, or in general long-range temporal correlations (LRTCs). Accordingly, slow autocorrelation decays have been observed using different experimental modalities, including studies on non-human primates \cite{Ogawa2010,Murray2014}, human electrocorticography \cite{Honey2012} as well as fMRI \cite{Stephens2013}.
In a large body of EEG research, slow timescales associated with long-range temporal correlations have been observed and studied in the amplitude fluctuations of ongoing neuronal oscillations \cite{Linkenkaer2001,Nikulin2004,Linkenkaer2007,Monto2007,Smit2011,Berthouze2010,Palva2013,Kantelhardt2015,Samek2016}.
Together, these studies provide strong support for the existence of slow timescales associated with LRTCs in cortical activity and their important role for the integration of information in brain networks. 

The link between LRTCs and information processing is further supported by the insight that LRTCs are generic features of a critical state \cite{Bak1987}. 
Critical dynamics has been shown to provide advantageous features for network computation and information processing \cite{Langton1990,Haldeman2005,Bertschinger2004,Shew2009}.
Numerous computational and experimental studies \cite{Bornholdt2000,Beggs2003,Haldeman2005,Meisel2009,Shew2009,Meisel2012,Markram2015} have provided support for the hypothesis that also brain network operate at a critical state and thereby take advantage of the computational capabilities provided by criticality. In this context, the observation of LRTCs in EEG dynamics has been taken as supporting evidence for the criticality hypothesis \cite{Linkenkaer2001}. 

In a previous study, we found various signatures of critical dynamics to fade over the course of sleep deprivation \cite{Meisel2013}. 
These observations led to the hypothesis \cite{Meisel2013} that deviations or disruptions of critical dynamics and its advantageous features for network computation might underlie the impaired cognitive functioning observed during sleep deprivation \cite{Banks2007}. The question whether LRTCs are similarly affected in a wake-time dependent manner which, however, is still open.

Here, we systematically characterize the LRTCs governing cortical dynamics during extended wake. We find that the initially slow timescales characterizing the autocorrelation decay in EEG amplitude fluctuations are progressively shortened as sleep deprivation progresses. This decline becomes evident when taking changes in EEG signal power into appropriate consideration. The results support a hypothesis on the network function of sleep, to re-organize cortical networks towards critical dynamics with long-range temporal memory for optimal function during wake.


\section*{Materials and Methods}
\subsection*{EEG recordings during prolonged wakefulness}

We analyzed wake electroencephalogram (EEG) recordings of eight healthy young right-handed males (23.0 $\pm$ 0.46 years; mean $\pm$ SEM) over the course of 40 hours of sustained wakefulness (data from a previous study; for details see \cite{Finelli2000}). Participants were under constant surveillance for the duration of the experiment. Waking EEG over the course of sleep deprivation was recorded every three hours over 14 sessions, starting at 07:00. Another waking EEG was recorded after a recovery night of sleep, totaling 15 EEG sessions in all. Sessions consisted of a 5 min eyes-open period, followed by a 4-5 min eyes-closed period and a final 5 min eyes-open period. Twenty-seven EEG derivations (extended 10-20 system; reference electrode 5 \% rostral to Cz) were sampled at 256 Hz (high-pass filter at 0.16 Hz; anti-aliasing low-pass filter at 70 Hz). 

\subsection*{Preprocessing of EEG signals}
Artifacts including eye blinks were marked by visual inspection. All analyses were performed on artifact-free signal segments during the eyes open condition. Electrodes were re-referenced to average reference.  Segments for further analysis were chosen to be 20 seconds long (5120 samples) as a balance between, on one hand, including many segments in the analysis and, on the other hand, having long enough segments to assess timescales related to long-range temporal correlations (LRTCs).

\subsection*{Estimation of long-range temporal correlations in EEG}

\subsubsection*{Autocorrelation}
Autocorrelation functions were dervied from the envelope of ongoing oscillations in EEG signals in confined frequency bands. Specifically, artifact-free EEG signals of 20-s duration were filtered in the respective frequency band (theta: 4-8 Hz; alpha: 8-12 Hz; beta/low gamma: 12-30 Hz; phase neutral filter by applying a third order butterworth filter in both directions) and the signal envelope was derived using the absolute value of the Hilbert transform. 
The autocorrelation function $ACF(s)$ of the envelope signal $x(t)$ with length $N$, mean $\mu$ and variance $v$  was then derived by

\begin{equation}
\label{ACF}
ACF(s)=\frac{\sum_{t=1}^{N-s} (x(t)-\mu)(x(t+s)-\mu)}{v},\quad s=1,...,N/2.
\end{equation}

We quantified the autocorrelation function decay by its value at lag one, i.e. $s=1$ in equation \ref{ACF}. As an alternative, we also quantified autocorrelation by the first lag-value where the autocorrelation function was equal or below $0.5$. Both ways of quantification revealed similar results. 

\subsubsection*{Detrended fluctuation analysis}
As a second approach to assess the autocorrelation structure in our data, we applied detrended fluctuation analysis (DFA). DFA has been applied numerous times in EEG to quantify LRTCs before, see \cite{Hardstone2012} for a review. For each artifact-free, filtered (theta: 4-8 Hz; alpha: 8-12 Hz; beta: 12-30 Hz; phase neutral filter by applying a third order butterworth filter in both directions), 20-second signal segment, we first extracted the absolute value of the signal's Hilbert transform. This provided the amplitude envelope of the signal $x$. Next, we determined the signal profile $Y(i)$ by subtracting the mean of the signal and computing its cumulative sum:

\begin{equation}
\label{profile}
Y(i) = \sum_{k=1}^{i} x_{k} - \langle{x}\rangle
\end{equation}

All subsequent steps were performed on the profile of the signal.

We defined a set $T$ of window sizes on a logarithmic scale. These window sizes were equally spaced between 16 sampling points (0.0625 s) and 4096 sampling points (16 s). This range is a compromise between computing the average fluctuation function for a given window size from many signal sub-segments on the one side, and having adequate samples in each subsegment over which to fit a least squares regression in the detrending step on the other side \cite{Hardstone2012}.

Next, the fluctuation function $F(t)$ for each window size $t \in T$ was found using the following three steps. First, the profile was split into $N$ non-overlapping sub-segments of length $t$. Second, we fit a polynomial of a given, fixed order to each sub-segment and subsequently subtracted the fit from that sub-segment. The detrended sub-segment $Y_{d}(t)$ of length $t$ is thus:

\begin{equation}
\label{detrended}
Y_{d}(t) = Y(t) - p_{v}(t)
\end{equation}

where $p_{v}(t)$ is the polynomial fit of the $v$th sub-segment. Third, the standard deviation was calculated for each detrended sub-segment. $F(t)$ for a given window size $t$ was then calculated as the average standard deviation across all sub-segments of size $t$.

Finally, $F(t)$ was plotted for all window sizes on log-log axes.  If a time series is characterized by long-range temporal correlations, then $F(t)$ increases with window size $t$ according to the power law:

\begin{equation}
\label{hurstExponent}
F(t) \propto t^{\alpha}
\end{equation}

Thus, from this log-log plot, the DFA scaling exponent (or Hurst exponent) can be estimated as the slope $\alpha$ via linear regression. $\alpha > 0.5$ indicates long-range correlations, $\alpha < 0.5$ indicates an anti-correlated signal and $\alpha = 0.5$ indicates an uncorrelated signal.

To accurately estimate the DFA scaling exponent, it is necessary to choose the proper range of window sizes over which to perform linear regression. Filtering can introduce short-range correlations and cause an overestimation of the exponent, if included \cite{Hardstone2012}. To determine the effect of our filters, we simulated a set of 100 white noise segments of the same length as our analyzed signal segments (20 s at 256 Hz sampling), applied each of our bandpass filters to the set, computed the average fluctuation function for each window size and plotted the log-log DFA plot. This kink visible in this plot demonstrates the impact of filters on our analysis (Fig. \ref{fig_2} c). To avoid the influence of filters when estimating the scaling exponent, we thus chose the fitting range sufficiently away from it between 2 seconds (512 sampling points) and 16 seconds (4096 sampling points; Fig. \ref{fig_2} c, Fit Range).

\subsection*{Signal power}
Power analyses were done on the same artifact-free 20-s segments used in the timescales analyses to assess LRTCs. For each frequency band of interest (4-8 Hz, 8-12 Hz, 12-30 Hz), EEG power density spectra were computed from filtered signals (phase neutral filter by applying a third order butterworth filter in both directions) in the respective frequency band (FFT routine; 4 s, non-overlapping Hanning window). The power value for each 20-s interval and channel was then obtained as the sum across the frequency band of interest.
Local slopes of the power spectral density were computed across lower frequencies in double-logarithmic coordinates.

\subsection*{Stastical Tests}
Changes in signal power, DFA scaling exponent and autocorrelation were tested for significance using ANOVA and post hoc t--tests. We also used linear regression based on least-squares fits.

\section*{Results}

\subsection*{Distinct changes in EEG frequency power and power spectrum slope during sleep deprivation}

Extended wakefulness is known to be associated with distinct changes in EEG power during waking and consecutive recovery sleep \cite{Torsvall1987,Cajochen1995,Aeschbach1999,Strijkstra2003}. In the wake EEG, the increase in slow frequency power in the theta range (4-8 Hz) is a particularly well-established sleep deprivation hallmark \cite{Finelli2000}. 
It has been pointed out that the estimation of long-range temporal correlations (LRTCs) in EEG can, in principle, be affected by EEG power  \cite{Linkenkaer2001,Linkenkaer2007,Monto2007,Smit2011}. We thus started by throughly assessing signal power and its changes over the course of sleep deprivation as this is of central importance for the following characterization of timescales and LRTCs.

We analyzed artifact-free EEG segments of 20-seconds duration during the eyes-open condition. 
In our data, we observed, in line with previous studies, an increasing signal power in theta as well as in higher frequency ranges including the beta/low gamma range (12-30 Hz; for simplicity referred to as beta in this manuscript) during sleep deprivation (Fig. \ref{fig_1}). The significance of power changes over the course of sleep deprivation was evident by linear regression (theta: $r=0.80$, $p<0.05$; beta: $r=0.62$, $p<0.05$) and the difference between beginning (0-6 h) and end (33-39 h) of sleep deprivation (theta: $p<0.05$; beta: $p<0.05$; two-sample t--test). Alpha power (8-12 Hz), conversely, exhibited the well-known circadian modulation but no significant cumulative power change during sleep deprivation (linear regression: $r=0.15$, $p=0.61$; 0-6 hours vs 33-39 hours: $p=0.46$; Fig. \ref{fig_1} b). Besides these changes in confined frequency bands, we also observed a more shallow slope of the power spectral density (PSD) across a range of lower frequencies after sleep deprivation. The estimation of local slopes from PSD showed that this change to a more shallow slope is significant, and is reversed following recovery sleep (Fig. \ref{fig_1} a). 

While a large portion of EEG analyses has traditionally focused on restricted frequency bands, the underlying scale-free nature of the EEG PSD has been shown to similarly reveal important information, e.g. with regard to task performance modulation \cite{Miller2009,He2010}.
From a dynamical perspective, the scale-free (or 1/f) scaling observed in the PSD may also indicate that the underlying system exhibits long-range spatio-temporal correlations \cite{Bak1987} and slow autocorrelation decay \cite{Talkner2000}. Interpreted within this framework, changes in the PSD slope, as observed in EEG during sleep deprivation here, could consequently be indicative of changes in the autocorrelation structure and long-range temporal correlations.
However, other mechanisms, such as passive filtering, have also been shown to be able to produce scale-free EEG power spectra, even in the absence of any long-range temporal correlations \cite{Bedard2006}. 
Methodologically, it can be difficult to obtain robust power spectrum estimates from short time series.
For these reasons, the structure of temporal correlations in EEG has been primarily studied in the amplitude envelope of filtered oscillations within confined frequency bands \cite{Linkenkaer2001,Berthouze2010,Hardstone2012,Palva2013,Samek2016}.

\subsection*{Decline in long-range temporal correlations during sleep deprivation}

To investigate the existence and the dynamics of long-range temporal correlations during sleep deprivation in our data, we calculated the amplitude envelope of ongoing oscillations in confined frequency bands from all channels of artifact-free EEG segments. We first focused on signal envelopes in the alpha (8-12 Hz) frequency band since average signal power between beginning and end of sleep deprivation did not change signifcantly there (Fig. \ref{fig_1} b). The impact of changes in signal power on LRTCs will be studied further below. 
The autocorrelation function exhibited a faster decay in recordings at the end of sleep deprivation compared to the beginning of sleep deprivation (Fig. \ref{fig_2} a). We quantified the decay by the autocorrelation value at time lag-1 which progressively decreased with increasing time awake and tended towards higher values after recovery sleep (Fig. \ref{fig_2} b). Both changes, the decline during sleep deprivation as well as the increase post recovery sleep, were statistically significant (0-6 hours vs 33-39 hours; 39 hours vs rec; $p<0.05$; two-sample t--test). 
An alternative quantification of the autocorrelation decay by monitoring the lag when autocorrelation values dropped below 0.5 for the first time exhibited similar results. 

As a second measure to investigate long-range temporal correlations during the course of sleep deprivation, we used detrended fluctuation analysis (DFA) \cite{Peng1994}. A substantial amount of research work has shown that DFA provides a robust measure for the autocorrelation structure of amplitude fluctuations in ongoing neurophysiological oscillations \cite{Linkenkaer2001,Nikulin2004,Monto2007,Poil2008,Berthouze2010,Palva2013,Samek2016}.
DFA relies on the quantification of signal fluctuations after removing their trend \cite{Hardstone2012}. It has been suggested that only a comparison of DFA results, using different detrending polynomials, yields full recognition of the trends \cite{Bashan2008,Kantelhardt2015}. We thus systematically applied DFA to the same amplitude time series of alpha oscillations as analyzed in the previous section using detrending polynomials from order 1 to 10. Our analysis again revealed a significant decrease of DFA scaling exponents for detrending orders 3 and higher (0-6 hours vs 33-39 hours; two-sample t--test; linear regression analysis; both $p<0.05$; Fig. \ref{fig_2} c, d). Since results did not change qualitatively when detrending was performed with higher order polynomials, we used detrending of 4th order ploynomials (DFA-4) for the remainder of the analyses. We observed DFA scaling exponents in the range of $>0.5$ to 1, which, in line with previous reports, is indicative of long-range temporal correlations in the EEG signals \cite{Linkenkaer2001}. Similar to the autocorrelation results we observed a decrease in scaling exponents during sleep deprivation indicative of a decline in long-range temporal correlations. The  potentially circadian component visible in DFA-4 on top of the decline became less prominent for higher orders of detrending while, at the same time, the DFA scaling exponent for the recovery point became higher, reaching a significantly higher value for detrending order 10 (39 hours vs rec; two-sample t--test; $p<0.05$).
Together, these results from the alpha band thus comprehensively indicate a decline in LRTCs with increasing time awake.

\subsection*{Influence of EEG signal power on long-range temporal correlation estimation}

The decline in timescales during sleep deprivation was observed in the alpha band where signal power did not change to overall higher or lower values as a function of time awake. The fact that power did not change is important since timescale estimates from autocorrelation and DFA can, in principle, be influenced by EEG signal power changes, as has been pointed out \cite{Linkenkaer2001,Linkenkaer2007,Monto2007,Smit2011}. The reasoning behind this argument is that the measured EEG signal always contains some noise component apart from the neuronal signal which, when EEG signal power is lower (higher) can shift estimates of LRTCs to lower (higher) values due to the relative contribution of noise in the signal. A low signal-to-noise ratio can thus lead to lower LRTC values due to the relative contribution of a fast decaying noise autocorrelation function \cite{Linkenkaer2007}.
In line with this conceptual argument, detailed analyses have shown that increasing amounts of noise (or, conversely, decreasing signal-to-noise ratios) can shift autocorrelation and DFA exponents to lower values and vice-versa in EEG \cite{Linkenkaer2007}.
Previous work aware of this issue has thus carefully investigated potential correlations between LRTC and power which were found to be marginal but significant (alpha and beta bands \cite{Linkenkaer2007,Smit2011}) to strong (theta \cite{Smit2011}).
Under conditions with drastic power changes in relevant frequency bands, such as during sleep deprivation in our data, it is therefore possible that these effects will impact the estimation of LRTC. 

In our data we observed that DFA scaling exponents were significantly correlated with signal power in all three frequency bands (Fig. \ref{fig_3} left column). In stark contrast to the alpha band, we observed an increase in the DFA scaling exponent in the theta band which closely mirrored the time course of theta power (Fig. \ref{fig_3} top). Specifically, the DFA exponent exhibited a circadian component on top of a monotonic increase which closely resembles the well-known increase of theta power during sleep deprivation \cite{Finelli2000}. This striking similarity in time course along with the significant correlation between power and scaling exponent thus suggest that the increase in DFA exponents in theta may be primarily due to the concomittant changes in signal power levels. This influence of signal power is also likely to determine the time course of DFA scaling exponents in the beta band to a large extent where signal power increases significantly too, and may similarly impact the signal-to-noise ratio and estimation of LRTCs \cite{Linkenkaer2007}. 

For better insight into the influence of the strongly increasing power on DFA exponents in theta and beta frequency bands, we repeated the analysis using only channels that did not show a significant change (0-6 hour vs 33-39 hour; two-sided t--test; $p>0.5$). Results were similar when different p-values (0.3-0.9) were used to exclude channels with significant power changes. By controlling for power changes (and in particular power increases) during sleep deprivation with this method, we observed no significant increases in DFA scaling exponents in both theta (7/189 remaining channels) and beta bands (31/189 channels remaining; 0-6 hours vs 33-39 hours; $p>0.05$; two-sample t--test; Fig. \ref{fig_3} right column). In fact, in theta band, scaling exponents even exhibited a slight decrease, albeit not reaching the significance level. When we controlled for power changes in the same way in the alpha band (87/189 channels remaining), the significant decrease in DFA scaling exponents during sleep deprivation persisted ($p<0.05$; two-sample t--test). Together, these analyses thus provide strong indication that the apparent increase of LRTC in theta and beta is due to the concomittant increase in signal power. Conversely, when no significant power changes are present, such as in the alpha band, then LRTCs decline during sleep deprivation. 

\subsection*{LRTC decline is predominantly found in frontal and parieto-occipital regions}
Our described trends so far were based on averages across all channels.
The decline in LRTC could also be observed at the individual channel level (Fig. \ref{fig_4} a; alpha band). The topographical distribution of sleep deprivation changes (0-6 hours vs 33-39 hours) showed that LRTC decreases are visible across all EEG channels and are particularly prominent over frontal and parieto-occipital leads for both autocorrelation and DFA (Fig. \ref{fig_4} b, c).

\section*{Discussion}
In the present work, we report the decline of long-range temporal correlations (LRTCs) in cortical activity during sustained wakefulness. We estimated the timescales of temporal correlations in EEG amplitude fluctuations of ongoing neuronal oscillations from the autocorrelation function directly as well as using detrended fluctuation analysis. While both analyses indicated the presence of long timescales associated with a slow autocorrelation decay at the beginning of sleep deprivation, sustained wakefulness led to a progressive decline with shorter timescales. Our results provide a novel perspective on the changes of cortical network dynamics with implications for their information processing capabilities during sustained wakefulness. They provide a missing link to previous findings indicating a disruption of critical dynamics during sleep deprivation \cite{Meisel2013}, and highlight the importance of adequately taking into consideration signal power changes when assessing long-range temporal correlations in EEG. 

Our main analyses focused on LRTCs in the alpha where signal power exhibits a circadian modulation, but no significant overall in- or decrease as a function of time awake. Under these conditions, a clear decline in LRTCs quantified by autocorrelation directly and by detrended fluctuation analysis could be observed during sustained wakefulness. Conversely, in the other frequency bands studied (theta and beta), signal power increased significantly over the course of sleep deprivation which compromised the estimation of LRTCs in these bands. The relevance of signal power for the estimation of LRTCs has been pointed out and studied several times before \cite{Linkenkaer2007,Monto2007,Smit2011}. Broadly speaking, estimates of LRTCs can be biased towards lower, more noise-like values when signal power (and thus the signal-to-noise ratio) is lower and vice-versa. Consequently, Linkenkaer-Hansen et al. observed positive, albeit weak, correlations between signal power and DFA exponents in some EEG channels \cite{Linkenkaer2007}.
In a similar vein, Smit et al. reported marginal correlations of LRTCs with power in the alpha and beta bands and substantial correlations in the theta band, which led the authors to conclude that the stronger signal-to-noise ratio in this frequency band may have caused increased LRTCs \cite{Smit2011}. 
Our results confirm this dependence and indicate that these effects are particularly important to consider in experimental conditions when signal power changes dramatically in some frequency bands, such as during sleep deprivation \cite{Finelli2000}. To make a robust conclusion about changes in LRTCs, or any other EEG measure for that matter, in an experiment, it is therefore important to carefully consider whether these changes could be caused by concomitant alterations in signal power. Here, we observed decreasing LRTCs during sleep deprivation when power remained constant (alpha band) and an apparent increase in the other bands (beta and theta) only when there was a drastic concomitant power increase. This is similar to our observations of synchronization measures in a previous study \cite{Meisel2013}, which were also found most predominant in the alpha band. Thus, these changes in LRTCs and synchronization measures cannot be explained as a direct consequence of the alterations in spectral power, which are most evident in the theta band \cite{Finelli2000}.

Differences in signal power and signal-to-noise ratio may potentially also help to reconcile some perhaps contradictory findings.
In a recent study, larger LRTCs associated with higher insomnia complaints were observed within groups of insomnia patients and control subjects, and interpreted as a sign of being closer to criticality \cite{Colombo2016}. Interestingly, no LRTC differences were observed between control and insomnia patient groups.
In light of the strong impact of signal power on estimates of LRTCs in our data, especially under conditions with high sleep pressure, it is possible that signal power changes associated with worse sleep quality may have contributed to the apparent increase in LRTCs in these subjects, similar to our data.

Long-range temporal correlations are generic features of systems in the vicinity of a critical state \cite{Bak1987}. 
The observation of LRTCs in EEG has thus been taken as additional evidence for a growing body of computational and experimental studies indicating that cortical neural networks operate at some sort of critical state \cite{Bornholdt2000,Linkenkaer2001,Beggs2003,Haldeman2005,Shew2009,Meisel2012,Markram2015}. 
Interpreted within this criticality framework, the decline of LRTCs during sustained wakefulness complements the observed fading of other signatures of critical dynamics during extended wake \cite{Meisel2013}.
This framework also provides an interesting link between network function and the often-observed impairments of cognitive capabilities during extended wake. 
Critical dynamics are often regarded to support optimal computational functioning \cite{Langton1990,Bertschinger2004,Haldeman2005,Kinouchi2006,Shew2009}. The decline of LRTCs and other signatures of critical dynamics during prolonged wakefulness suggests that the brain can benefit less from the computational advantages of critical or near-critical dynamics. Behavioral observations of impaired cognitive functioning and information processing after sleep deprivation \cite{Banks2007} might be the result of these functional deficits.

With regard to the mechanisms responsible for the observed disruption of long timescales, it is conceivable that the control of cortex dynamics through subcortical regions might play a role.
It is known that several subcortical structures in the brainstem, hypothalamus and basal forebrain regulate the maintenance of waking and sleep states through neuro-modulatory action \cite{Jones2005}. During extended wake, these subcortical structures could transiently disrupt the ongoing cortical dynamics leading to an apparent decline in long-range temporal correlations. 
Alternatively, it is conceivable that sleep/wake dependent structural changes in cortical networks, such as changes in average synaptic strenghts \cite{Tononi2014,DeVivo2017}, might drive network dynamics closer/further away from criticality and thereby effect cortical timescales.
Independent of the underlying mechanisms, our results indicate that sleep is important to reorganize cortical networks towards dynamics with long-range temporal correlations for optimal function during wake.


\clearpage
\section*{Author Contribution}
CM conceived the analysis. CM and KB analyzed the data and wrote the manuscript. PA and DP contributed data and provided critical revision of the article. All authors reviewed the manuscript.

\section*{Acknowledgements}
This study was supported by the Intramural Research Program of the NIMH.  This study utilized the high-performance computational capabilities of the Biowulf Linux cluster at the National Institutes of Health, Bethesda, Md. (http://biowulf.nih.gov).

\section*{Additional Information}
The authors declare no competing financial interests.

\newpage
\begin{figure}
\centering
\caption{Changes in EEG power during sleep deprivation. a, EEG power density at the beginning of sleep deprivation (0-6 hours, blue) and end of sleep deprivation (33-39 hours, red; grand average across 8 subjects and all channels, error bars indicate s.e.m.). Local slopes of the PSD were obtained in the ranges indicated by the solid lines. Right: Local slopes indicate a more shallow PSD at the end of sleep deprivation compared to beginning of sleep deprivation and after recovery sleep. b, Power changes in frequency bands. Straight black lines indicate linear regression results; bars indicate the average power values at the beginning (0-6 hours) and at the end (33-39 hours) of sleep deprivation. Of the three frequency bands, only the alpha band (8-12 Hz) showed no significant change during sleep deprivation as judged by linear regression and difference between bars ($\Delta \rm{Power}$). \label{fig_1}
}
\end{figure}

\begin{figure}
\centering
\caption{Decline in signal autocorrelation and long-range temporal correlations during sleep deprivation. a, Autocorrelation function of the signal envelope in the alpha band (0-6 hours, blue; 33-39 hours, red; inset, autocorrelation functions until 1s). b, Faster autocorrelation decay during sleep deprivation (0-39 hours of sleep deprivation and after consecutive recovery sleep, rec). The solid black line corresponds to the mean across all channels from all 8 subjects, error bars indicate to s.e.m. c, Detrended fluctuation analysis of the of the signal envelope in the alpha band. Surrogate white-noise data are depicted in grey. DFA scaling exponents were obtained from linear fits in double-logarithmic coordinates aver the depicted fitting range. d, Decline in long-range temporal correlations estimated by the DFA scaling exponent. The inset shows results for different polynom detrending orders. Results settle for detrending order of three and higher as judges by linear regression analysis and the difference between early (0-6 hours) and late (33-39 hours) of sleep deprivation ($\Delta \rm{DFA}$). \label{fig_2}
}
\end{figure}

\begin{figure}
\centering
\caption{Influence of signal power on estimating long-range temporal correlations. DFA scaling exponents exhibited a significant correlation with signal power in all frequency bands investigated (left column). Middle column: DFA scaling exponents (black) follow a very similar trajectory to signal power (green) particularly when power is changing significantly, as is the case in the theta (top) and beta (bottom) bands. Right column: DFA increases between early (0-6 hours, blue) and late (33-39 hours, red) sleep deprivation vanish in the theta and beta bands when only channels with no significant power change (power controlled channels) are considered for DFA estimation. Conversely, the decline in DFA scaling exponents in the alpha band remains. This suggests that it is the strongly increasing signal power in theta and beta bands that leads to an apparent increase of DFA scaling exponents in these frequency bands. \label{fig_3}
}
\end{figure}

\begin{figure}
\centering
\caption{The decline of timescales during sleep deprivation occurs over broad cortex areas with a particular dominance in frontal and parieto-occipital regions. All plots correspond to results from the alpha band (8-12 Hz). a, Exemplary time courses of from two individual channels indicating a strong decline in autocorrelation values during sleep deprivation. b, Topographic distribution of changes in lag-1 autocorrelation during sleep deprivation (difference between average values from 0-6 hours and average values from 33-39 hours). Each marker corresponds to one EEG channel. While a decrease in autocorrelation is evident in all channels, changes a praticularly prominent in frontal and parieto-occipital regions. c, Topographic distribution of sleep deprivation changes in DFA scaling exponents. d, Topographic distribution of average power in the alpha band.
\label{fig_4}
}
\end{figure}


\beginsupplement

\clearpage

\end{document}